\documentclass[aps,prd,a4paper,showpacs,showkeys,preprint,nofootinbib]{revtex4-1}

\usepackage{amssymb}
\usepackage{amsmath}
\usepackage{bm}
\newcommand{\christ}[2]{{#1 \brace #2}}

\newcommand{\diag}{\ensuremath{\mathrm{diag}}}

\newcommand{\qo}[1]{\textquotedblleft #1\textquotedblright}
\begin{document}
\title{Weyl-gauge invariant proof of the Spin-Statistics Theorem}
\author{Enrico Santamato}
\affiliation{Dipartimento di Fisica, Universit\`{a} di Napoli Federico II, Complesso Universitario di Monte S. Angelo, 80126 Napoli, Italy}
\email{enrico.santamato@na.infn.it}
\author{Francesco De Martini}
\affiliation{Accademia Nazionale dei Lincei, via della Lungara 10, 00165 Roma, Italy}
\email{francesco.demartini@uniroma1.it}
\begin{abstract}
The traditional standard theory of quantum mechanics is unable to solve the spin-statistics problem, i.e. to justify the utterly important \qo{Pauli Exclusion Principle} but by the  adoption of the complex standard relativistic quantum field theory. In a recent paper (Ref.~\cite{SantamatoDeMartini2015}) we presented a proof of the spin-statistics problem in the nonrelativistic approximation on the basis of the \qo{Conformal Quantum Geometrodynamics}. In the present paper, by the same theory the proof of the Spin-Statistics Theorem is extended to the relativistic domain in the general scenario of curved spacetime. The relativistic approach allows to formulate a manifestly step-by-step Weyl gauge invariant theory and to emphasize some fundamental aspects of group theory in the demonstration. No relativistic quantum field operators are used and the particle exchange properties are drawn from the conservation of the intrinsic helicity of elementary particles. It is therefore this property, not considered in the Standard Quantum Mechanics, which determines the correct spin-statistics connection observed in Nature~\cite{SantamatoDeMartini2015}. The present proof of the Spin-Statistics Theorem is simpler than the one presented in Ref.~\cite{SantamatoDeMartini2015}, because it is based on symmetry group considerations only, without having recourse to frames attached to the particles.
\end{abstract}
\maketitle
\section{Introduction}\label{sec:intro}
Within the domain of the Standard Quantum Mechanics (SQM) the Spin-Statistics Theorem (SST) could not be demonstrated but by making use of the full complexity of the quantum field theory, e.g. the one accounted for by the well known book by Streater and Wightman~\cite{StreaterWightman2000}. As a consequence of this complexity the SST is generally taken as a postulate which is indeed a \qo{Principle}: the celebrated \qo{Pauli exclusion principle}. Pauli's derivation of the Spin-Statistics connection is based on the assumption that quantum fields operators either commute or anticommute for space-like separation, on the positive metric of the Hilbert space and on the impossibility of negative energy states. The initial approach by Fiertz~\cite{Fierz1939} and Pauli~\cite{Pauli1940} was then refined and generalized by Schwinger~\cite{Schwinger1958}, Wightman~\cite{Wightman1956}, Duck et. al.~\cite{DuckSudarshanWightman1999}, and others (see~\cite{DuckSudarshan1998a,Romer2002} and references therein). However, even after so many years, today the existing proof of SST is very formal and the underlying physics is obscure. Accordingly, we may agree with Duck and Sudarshan by saying that \qo{everyone \textit{knows} the Spin-Statistics Theorem, but no one \textit{understands} it}~\cite{DuckSudarshan1998}. Indeed, the SST main statement (\qo{no more than one fermion can occupy a given quantum state, while any number of bosons can occupy the same quantum state}) is so simple and the theorem so fundamental that it has always been taken as a surprising fact that the foundations of SST rely in the deep arguments of relativistic quantum field theory. As Feynman said: \qo{\textit{this probably means that we do not have a complete understanding of the fundamental principle involved}}~\cite{FeynmanLeighton2011}. In the last decades several publications appeared aimed at finding more acceptable derivations of the Pauli principle, including attempts in the settings of nonrelativistic quantum mechanics, but none of the arguments put forward consists a satisfactory elementary proofs of the SST: in facts they leave the situation essentially as Feynman found it 30 years earlier~\cite{DuckSudarshanWightman1999,FeynmanLeighton2011}.\\

In a recent paper~\cite{SantamatoDeMartini2015}), we presented a demonstration of the SST in the framework of the Conformal Quantum Geometrodynamics (CQG)~\cite{SantamatoDeMartini2013a,SantamatoDeMartini2013,SantamatoDeMartini2012,Santamato2008,DeMartiniSantamato2015,DeMartiniSantamato2014b, DeMartiniSantamato2014,DeMartiniSantamato2013} in the nonrelativistic approximation. Here we extend the proof to the relativistic case. Although the proof of the SST runs similarly in the relativistic domain, this one allows to clarify better some geometric and group theoretical aspects of the \textit{intrinsic helicity} of quantum particles, a quantity not considered by the SQM theory of the \qo{spin}\cite{DeMartiniSantamato2014a}, which plays nevertheless a deep role in our proof of the SST.\\

The CQG provides a realistic and complete approach to Quantum Mechanics with no adoption of the quantum \qo{wave functions} as the primary object of the theory. In the CQG, the quantum features arise from the requirement of invariance of the theory under local change of the metric \textit{calibration} [see Sec.~\ref{sec:CQG}]. This invariance principle is due to Weyl~\cite{Weyl1952} and extends the well known Einstein's general relativity principle to the following:  \qo{\textit{all physical laws should be invariant under arbitrary change of coordinates and arbitrary local change of metric calibration}}. As shown by Weyl~\cite{Weyl1952} the requirement of calibration invariance of physical laws implies that in a parallel transport the length $\ell=g_{ij}a^ia^j$ of a contravariant vector $a^i$ $(i=1,...\dots,n)$ changes  as $\delta\ell=-2\ell d\phi$, where $d\phi=\phi_idq^i$ is a differential form in the coordinates $q^i$ spanning the considered $n$-dimensional manifold. The Weyl connections associated with this parallel transport are
\begin{equation}\label{eq:Gammaijk}
   \Gamma^i_{jk} = \christ{jk}{i}+\delta^i_j\phi_k+\delta^i_k\phi_j-g_{jk}\phi^i
\end{equation}
where $\christ{jk}{i}$ are the Christoffel symbols of the metric $g_{ij}$ and $\phi^i=g^{il}\phi_l$. The Weyl connections (\ref{eq:Gammaijk}) are  invariant under the Weyl gauge~\cite{Weyl1952} ($\partial_i$ denotes the ordinary partial derivative with respect to the coordinate $q^i$)
\begin{subequations}\label{eq:gauge}
\begin{eqnarray}
   g_{ij} \to \lambda^2(q)g_{ij} \label{eq:gijgauge}\\
   \phi_i \to \phi _i - \frac{\partial_i\lambda}{\lambda}\label{eq:phigauge}
\end{eqnarray}
\end{subequations}
and so are the curvature tensors $R^i_{jkl}$ and its contractions $R_{ij}=R^k_{ikj}$, $W_{ij}=R^k_{kij}=\partial_i\phi_j-\partial_j\phi_i$~\footnote{In some axiomatic approaches to the structure of spacetime, the so-called \qo{Weyl tensor} $W^i_{jkl}$ is used in place of the curvature tensor $R^i_{jkl}$. The Weyl tensor is obtained from the full curvature tensor by subtracting out various traces and describes the tidal part of the gravitational forces. The Weyl tensor is unrelated to the scalar curvature $R_W$ in Eq.~(\ref{eq:RW}), because all contractions of $W^i_{jkl}$ are zero.}. From the connections (\ref{eq:Gammaijk}) one obtains the Weyl scalar curvature $R_W=g^{ij}R_{ij}$~\cite{Weyl1952}
\begin{equation}\label{eq:RW}
  R_W = R_g - (n-1)[(n-2)g^{ij}\phi_i\phi_j+2g^{ij}\nabla_i\phi_j]
\end{equation}
where $R_g$ and $\nabla_i$ denote the Riemann scalar curvature and the covariant derivative with respect to the coordinate $q^i$ calculated from the Christoffel symbols of the metric tensor $g_{ij}$, respectively. We see that the requirement of calibration invariance induces a curvature in the manifold even if the Christoffel symbols vanish as in the case of Euclidean or Minkowski metric. The calibration invariance of physical laws is equivalent to the requirement of invariance of all equations of the theory under the Weyl gauge transformations (\ref{eq:gauge}). Tensors $T(q)$ which under the Weyl gauge (\ref{eq:gauge}) transform as $T(q) \to \lambda^{w(T)}(q)T(q)$ are said to have Weyl weight $w(T)$. For example, Eq.~(\ref{eq:gijgauge}) shows that $w(g_{ij})=2$ and $w(g^{ij})=-2$. The weight of $R_W$ is $w(R_W)=-2$. For such tensors we may introduce the Weyl co-covariant derivative $D_i T=\nabla^{(W)}_iT+w(T)\phi_i$, where $\nabla^{(W)}_i$ is the covariant derivative with respect to the Weyl connections (\ref{eq:Gammaijk}). The main properties of the co-covariant derivative is that the metric tensor is constant ($D_i g_{jk}=0$ ) and that leaves the Weyl weight of $T$ unchanged ($w(D_iT)=w(T)$).\\

The similarity of the gauge transformation (\ref{eq:phigauge}) with the electromagnetic gauge transformations led Weyl himself to his well known attempt to unify electromagnetism and gravity in a single geometric framework~\cite{Weyl1918,Weyl1952}. Despite the well known Einstein's criticism~\cite{einstein1918} and the fact that electromagnetism has been unified successively with the weak interactions, many investigations appeared to explore the potentiality of Weyl's generalization of Riemannian spacetime~\cite{Dirac1973,Utiyama1973,Utiyama1975,Cheng1988,HayashiKasuyaShirafuji1977,Wheeler1990,Wheeler1998}. We notice also the remarkable fact that a number of axiomatic approaches for deducing spacetime structure from basic concepts such as light rays and freely falling test particles all end by assigning a Weylian, not Riemannian, structure to spacetime~\cite{J.EhlersA.Schild1972,HochbergPlunien1991,faraoni_beyond_2011,Trautman2012,FatibeneGarrutoPolistina2015}. All works quoted above apply Weyl's geometry to extended gravity in the four dimensional spacetime. The proposal of the CQG is to apply Weyl's geometry to quantum mechanics, instead, leaving gravitation fixed in the background. This possibility has been conjectured some time ago~\cite{Santamato1984,Santamato1984a,Santamato1985,Santamato1988} after noticing that the second term on the right of Eq.~(\ref{eq:RW}) for the scalar Weyl curvature $R_W$ can be reduced to the well known \qo{quantum potential} appearing in Bohm's approach to quantum mechanics~\cite{Bohm1952,Bohm1952a,BohmHiley1995}. These earlier works consider the fields defined in spacetime only and do not account for what may be considered the quantum feature par excellence of elementary particles: i.e. the Spin. Before concluding this Section, we present here a brief outline of the CQG applied to a quantum elementary particle.  More details will be provided in the rest of the paper.\\

To describe a quantum particle, the CQG extends the 4-D spacetime $M_4$ to the larger $n$-dimensional space $V_n=M_4\times G$, where $G$ is the $K$-dimensional parameter space of suitable group $G$ used to describe the internal symmetry of the particle. In the case of a particle with spin, we have $G=\tilde L_+$, the proper orthochronous Lorentz' group. The coordinates $q^i$ $(i=1,\dots,n)$ $(n=4+K)$ can be grouped into $q^i=(x^\mu,y^\alpha)$, where $x^\mu$ $(\mu=0,1,2,3)$ are the spacetime coordinates and $y^\alpha$ $(\alpha=1,\dots,K)$ are the parameters of $G$. The metric tensor of $V_n$ is taken as the direct sum $g_{ij}=g_{\mu\nu}\oplus g_{\alpha\beta}$. The CQG is based on an action integral for two scalar fields $\rho$ and $\sigma$ defined in $V_n$ (the Lagrangian is given in Eq.~(\ref{eq:L}) below). Palatini's variational method is applied to obtain the field equations of $\rho$ and $\sigma$ as well as the affine connections $\Gamma^i_{jk}$ $(i,j,k=1,\dots,n)$. The affine connections are found to be the Weyl connections (\ref{eq:Gammaijk}) with integrable Weyl field $\phi_i$. Therefore the geometry of $V_n$ is conformally Riemannian with scalar curvature given by Eq.~(\ref{eq:RW}). The field equations of $\rho$ and $\sigma$ are covariant with respect to the Weyl gauge (\ref{eq:gauge}) and are the fundamental equations of the CQG. A logarithmic \textit{ansatz} allows to group the coupled nonlinear equations for $\rho$ and $\sigma$ into a single linear wave equation for a complex scalar field $\Psi$. This reduction is made possible because the second term in the Weyl curvature (\ref{eq:RW}) is found to have the form of the well known bohmian \qo{quantum potential}~\cite{Bohm1952,Bohm1952a,BohmHiley1995}. The wave equation obeyed by $\Psi$ is not a known wave equation of the SQM and $|\Psi|^2$ has no definite physical meaning because it is not Weyl invariant. We must then find the mathematical and epistemological connection between the CQG and the SQM. This can be accomplished as follows. To each single valued solution $(\rho,\sigma)$ to the fundamental equations of the CQG is associated a Weyl invariant geometric structure formed by a bundle of trajectories in $V_n$ intersected by the surfaces $\sigma=\mathrm{\ const.}$ and by a Weyl invariant current density along these trajectories. We will call this geometric structure the \qo{complete figure} of the CQG. The existence of the complete figure makes the CQG compatible with the familiar statistical interpretation of the SQM and provides a link with quantum experiments based on particle count rates. The count rates foreseen by the CQG reproduce the results of the SQM, including quantum correlations violating Bell's inequalities, thus giving a solution to EPR paradoxes. The Weyl invariance of the complete figure ensures that the experimental outcomes have a definite physical meaning. The epistemological equivalence between the CQG and the SQM is thus established. The mathematical equivalence of the CQG with the SQM is accomplished by expanding the scalar wave function $\Psi$ along the harmonics (i.e. the irreducible representations) of the internal group $G$ with coefficients $\psi^q(x)$ depending on the spacetime coordinates only. The fields $\psi^q(x)$, however, cannot still be identified with the spinor wave functions of the SQM. A complete mathematical equivalence between the CQG and the SQM is obtained only after having observed that the fundamental equations of the CQG foresee a conserved fundamental quantity which characterizes the elementary particles: the \qo{intrinsic helicity}. The intrinsic helicity can be considered a fixed property of elementary particles akin mass and plays a fundamental role in our proof of the SST. The SQM ignores the intrinsic helicity which may explain why the SQM is unable to obtain a simple proof of Pauli's principle. The conservation of the intrinsic helicity is not compatible with the expansion of $\Psi$ in terms of the harmonics of the whole internal group $G$. The expansion must then be restricted, in general, over the harmonics of suitable homogeneous quotient space $G/H$, with $H$ subgroup of $G$. The coefficients $\psi^q(x)$ of the expansion of the scalar wave function $\Psi$ along $G/H$ can finally be identified with the spinor fields used in the SQM. In the case of the particle with spin, we have $G/H = \tilde L_+/R_2$, where $R_2$ is the group of 2D rotations around a fixed axis. We may think $R_2$ as representing the rotation of the particle \qo{around itself}. The harmonic expansion along $\tilde L_+/R_2$ restricts the allowed values $s$ of the intrinsic helicity of the spinning particle to integer and half integer numbers for bosons and fermions, respectively. The coefficients $\psi^q(x)$ of the expansion in the harmonics of $\tilde L_+/R_2$ are found to be Dirac's four components spinors which obey Dirac's equation.\\

Most of the claims quoted above have been demonstrated in previous works~\cite{SantamatoDeMartini2013a,SantamatoDeMartini2013,SantamatoDeMartini2012,Santamato2008,DeMartiniSantamato2015,DeMartiniSantamato2014b, DeMartiniSantamato2014,DeMartiniSantamato2013} and will not be discussed here. We conclude this Section by observing that the conservation of the particle intrinsic helicity and the consequent expansion of the scalar wave function along $\tilde L_+/R_2$ introduces a global phase factor $e^{is\gamma}$ to $\Psi$. As we will show in Sec.~\ref{sec:RSST}, when two identical particles are considered, this phase factor -- ignored in the SQM -- provides the right factor $(-1)^{2s}$ which originates Pauli's exclusion principle. The paper is organized as follows. In Sec.~\ref{sec:CQG} the action principle and the main equations of the CQG are presented. In Sec.~\ref{sec:stat} we discuss the complete figure and the statistical interpretation of the CQG. In Sec.~\ref{sec:onespin} we present the CQG approach to a single particle with spin. In Sec.~\ref{sec:RSST} we give our proof of the SST and, finally, in Sec.~\ref{sec:concl} our conclusions are drawn.
\section{The Conformal Quantum Geometrodynamics}\label{sec:CQG}
The CQG is based on a Weyl invariant action principle with a Weyl invariant Lagrangian density $L\sqrt{g}$ defined on the $n$-dimensional configuration space $V_n$ of the system spanned by coordinates $q^i$. Because $w(\sqrt{g})=n$, we must have $w(L)=-n$. The familiar choice $L=R_W$ is then inadequate, because $w(R_W)=-2$. If we consider only the four dimensions of spacetime a possible choice is $L=R_W^2$ as Weyl did~\cite{Weyl1952}. However, to have connection with Einstein's gravity, Weyl was forced to fix the gauge so to have $R_W = \mathrm{const.}=G$ with $G$ Newton's constant. A substantial improvement has been offered by Dirac~\cite{Dirac1973} by introducing an invariant constraint so to have $L = \Phi^2 R_W$ with Lagrange multiplier $\Phi$ of weight $w(\Phi)=-1$ in four dimensions. Dirac's Lagrangian was used also by Hochberg and Plunien~\cite{HochbergPlunien1991} who included the kinetic term $W_{\mu\nu}W^{\mu\nu}$ ($w_{\mu\nu}=\partial_\mu\phi_\nu-\partial_\nu\phi_\mu$, with $\phi_\mu$ Weyl vector)) and a scalar potential $V(\Phi)=\Phi^4$.\\

The Lagrangian used in the CQG is
\begin{equation}\label{eq:L}
   L = \rho [R + (\nabla_k\sigma- a_k)(\nabla^k\sigma-a^k)]
\end{equation}
where $R$ is the scalar curvature of the configuration space $V_n$ with metric tensor $g_{ij}$ and generic connections $\Gamma^i_{jk}$. The fields $a_i(q)$ represent externally applied fields which are prescribed. As said in the Introduction we assume $g_{ij}$ fixed (e.g. by external gravitational bodies) and take the variation of the action principle $\int L \sqrt{g}d^nq$ with respect to the connections $\Gamma^i_{jk}$ and the fields $\rho$ and $\sigma$. The variation of the action with respect the connections yields the Weyl connections (\ref{eq:Gammaijk}) with Weyl vector
\begin{equation}\label{eq:phi}
  \phi_i = \frac{1}{n-2}\frac{\partial_i\rho}{\rho}
\end{equation}
We see that the Weyl connection obtained from the variational principle is integrable with $\phi=\frac{1}{n-2}\ln\rho$ as Weyl potential. We have therefore  $W_{ij}=\partial_i\phi_j-\partial_j\phi_i=0$ so that the kinetic term is absent and the geometry is Riemannian up to conformal changes of the metric tensor. Once the variation was made, we may replace $R$ in the Lagrangian (\ref{eq:L}) with Weyl's scalar curvature $R_W$ so that the first term on the right of $L$ reduces to Dirac-Weyl's curvature Lagrangian with $\Phi^2=\rho$. Moreover, insertion of Eq.~(\ref{eq:phi}) into Eq.~(\ref{eq:RW}) yields
\begin{equation}\label{eq:RWrho}
   R_W = R_g +\left(\frac{n-1}{n+2}\right)\left[\frac{\nabla_k\rho\nabla^k\rho}{\rho^2}-2\frac{\nabla_k\nabla^k\rho}{\rho}\right]
\end{equation}
where $R_g$ is the Riemann scalar curvature built from the metric tensor $g_{ij}$. The variations with respect $\rho$ and $\sigma$ yield, respectively,
\begin{equation}\label{eq:HJ}
  (\nabla_k S- A_k)(\nabla^k S-A^k)+ \hbar^2\xi^2R_W = 0
\end{equation}
and
\begin{equation}\label{eq:cont}
   \nabla_k[\rho(\nabla^k S-A^k)] = 0
\end{equation}
with $\xi=\sqrt{\frac{n-2}{4(n-1)}}$ and $R_W$ given by Eq.~(\ref{eq:RWrho}). Weyl covariance is ensured by taking $w(\rho)=-(n-2)$ and $w(\sigma)=w(a_i)=0$. In Eqs.~(\ref{eq:HJ}) and (\ref{eq:cont}) we passed to dimensional units by setting $S=\xi\hbar\sigma$ and $A_i=\xi\hbar a_i$. Once the curvature (\ref{eq:RWrho}) is inserted into Eq.~(\ref{eq:HJ}), Eqs.~(\ref{eq:HJ}) and (\ref{eq:cont}) form a set of two coupled  nonlinear second-order partial differential equations for the scalar fields $\rho$ and $\sigma$, respectively. Note that Eq.~(\ref{eq:HJ}), regarded as an equation for $\sigma$, has the form of the Hamilton-Jacobi equation of the classical mechanics and that Eq.~(\ref{eq:cont}) regarded as an equation for $\rho$ has the form of a continuity equation for the Weyl-invariant current density
\begin{equation}\label{eq:curr}
   J^i=\rho(\nabla^i S -A^i)\sqrt{g}.
\end{equation}
We observe that the last term on the right of Eq.~(\ref{eq:RW}) is proportional to the \qo{quantum potential} used by Bohm in its \qo{ontological} approach to the SQM~\cite{Bohm1952,Bohm1952a,BohmHiley1995}. This suggests a possible relationship between Weyl's geometry and quantum mechanics. On the other hand, the first term on the right of $R_W$ is the Riemann curvature which depends on the metric tensor and its derivatives only. Thus, gravitational and quantum phenomena appear together on different footings: the metric tensor and the Riemann curvature $R_g$ in Eq.~(\ref{eq:RW}) are related to the gravitational forces, while the Weyl potential and the last term in the scalar curvature $R_W$ are related to quantum \qo{Bohmian} forces. We are implicitly assuming here that a gauge exists where this separation is meaningful. It should be noticed, however, that in other gauges gravitational and quantum forces are mixed. For example, in the gauge $\rho =$~const. where the geometry is purely Riemannian, gravitational and quantum forces are both encoded in the metric tensor. The Bohmian form of the \qo{quantum potential} in the Weyl scalar curvature $R_W$ suggests that a solution of the nonlinear problem posed by Eqs.~(\ref{eq:RW}), (\ref{eq:HJ}) and (\ref{eq:cont}) is conveniently attained by use of the \textit{ansatz}
\begin{equation}\label{eq:ansatz}
   \Psi=\sqrt{\rho}e^{iS/\hbar}
\end{equation}
by which the coupled Eqs.~(\ref{eq:HJ}) and (\ref{eq:cont}) are transformed into a single linear partial differential equation for the \textit{complex} field $\Psi$, viz.
\begin{equation}\label{eq:wave}
   (\nabla_k-A_k)(\nabla^k-A^k)\Psi-\hbar^2\xi^2R_g\Psi  = 0
\end{equation}
This equation has the form of the Klein-Gordon equation in curvilinear coordinates with the mass replaced by the Riemann curvature field $R_g$. The presence of the Riemann curvature is necessary for the Weyl-gauge invariance of Eq.~(\ref{eq:wave}). We will call $\Psi$ the \qo{scalar wavefunction} of the CQG. Solving Eq.~(\ref{eq:wave}) is equivalent to solve Eqs.~(\ref{eq:HJ}) and (\ref{eq:cont}) for $\rho$ and $S$. We notice, however, that because $\arg(\Psi) = S/\hbar + 2k\pi$ ($k$ integer), the scalar wavefunction $\Psi$ determines $S/\hbar$ only $\mod(2\pi)$. Then, once $\Psi$ is found, the right value of $k$ must be selected in order to satisfy the known global properties of $S$, as for instance singlevaluedness. In the following, the $\mod(2\pi)$ indetermination of the phase of $\Psi$ will be understood albeit not expressed explicitly.\\
\subsection{The statistical interpretation of the CQG}\label{sec:stat}
The CQG is an essentially geometric theory. Nevertheless, as expected from the SQM, a statistical interpretation can be also envisaged and this statistical interpretation has a Weyl invariant character. As observed above, Eq.~(\ref{eq:HJ}) has the form of the Hamilton-Jacobi equation of the classical mechanics. We may then associate to each single valued solution $S(q)$ of this equation a Carath\'{e}odory's \textit{complete figure} formed by a canonical bundle of trajectories $q^i(\tau)$ ($\tau$ is an arbitrary parameter along the trajectory) belonging to the family of surfaces $S(q) = \mathrm{const.}$~\cite{Rund1973}. Each curve of the canonical bundle obeys the Euler-Lagrange equations of the parameter invariant homogeneous Lagrange function $L^*(q,\dot{q}) = \sqrt{-R_W(q)g_{ij}(q)\dot{q}^i\dot{q}^j} + A_i(q)\dot{q}^i$ and, together with the conjugate momenta obtained by equating  $p_i = \partial L^*(q,\dot{q})/\partial \dot{q}^i$ with the gradient of $S$ (i. e. $p_i = \partial_iS$), forms a corresponding trajectory in the phase space. Moreover, if $C:q^i(\tau)$ is a trajectory belonging to the canonical bundle, we have
\begin{equation}\label{eq:DeltaS}
 S(P_2)-S(P_1) = \int_{P_1}^{P_2} L^*(q(\tau),\dot{q}(\tau))d\tau.
\end{equation}
Because $L^*$ is Weyl invariant, the complete figure is a Weyl invariant geometric object. The complete figure together with the current density $J^i$ given by Eq.~(\ref{eq:curr}) provides a Weyl invariant structure that allows a statistical interpretation of the CQG according to the SQM~\cite{Ballentine1970,Santamato1984a}. In particular, the current density $J_i$ provides a direct link with quantum measurement based on particle count rates. When two spin 1/2 particles in entangled state are considered, $J_i$ accounts for nonlocal EPR correlations too~\cite{SantamatoDeMartini2012,DeMartiniSantamato2012,DeMartiniSantamato2012a,DeMartiniSantamato2013,SantamatoDeMartini2013a,DeMartiniSantamato2014,DeMartiniSantamato2014b,DeMartiniSantamato2015}. The complete figure provides also a powerful tool to study the global properties of the quantum wavefunctions and, in particular, to prove the SST. We will consider therefore the existence of the complete figure as an essential feature of the CQG, which implies singlevaluedness of the function $S(q)$ in the considered region of the configuration space $V_n$.
\subsection{The connection between the CQG and the SQM}\label{sec:CQGtoSQM}
Equation~(\ref{eq:wave}) is not a known equation of the SQM. We need therefore to find how to connect Eq.~(\ref{eq:wave}) with the SQM formalism. We assume that the configuration space $V^N$ of a system of $N$ identical particles is the direct product $V^N=V\times\dots\times V$ ($N$ times) of the configuration space $V$ of one particle and, in turn, that $V$ is the direct product $V=M_4\times G$ of the 4-dimensional spacetime $M_4$ and the homogeneous $K$-dimensional parameter space of some group $G$, representing the internal symmetries of the particle. We have therefore $V^N=M^N_4\times G^N$, with $M_4^N=M_4\times\dots\times M_4$ and $G^N=G\times\dots\times G$. The coordinates $q^i$ of $V^N$ can be grouped as $q=(x,y)$, where $x$ collects the spacetime coordinates $x^\mu_h$ $(\mu=0,1,2,3)$ of the $N$ particles labeled by $(h=1,\dots,N)$ and $y$ collects the coordinates $y^\alpha_h$ $(\alpha = 1,\dots,K)$ of the internal symmetry group $G^N$. The fields $A_i(q)=A_i(x,y)$ $(i=1,\dots,N(4+K))$ in Eqs.~(\ref{eq:HJ}), (\ref{eq:cont}) and (\ref{eq:wave}) are assumed to be given by the direct sum $A_i(x,y) = \bigoplus_{h=1}^N A_h$, where
\begin{equation}\label{eq:Ah}
  A_h=(e/c)(A_\mu(x_h),K_\alpha^{\;\;m}(y_h)F_m(x_h))
\end{equation}
are vectors in the $(4+K)$-dimensional space $V$ of the single particle. In Eq.~(\ref{eq:Ah}) $A_\mu(x_h)$ and $F_m(x_h)$ are respectively the externally applied potentials and fields evaluated at the spacetime position $x_h$ of the particle $h$ and $K_\alpha^{\;\;m}(y_h)$ are the covariant Killing vectors of the symmetry group $G$ evaluated at $y_h$. Analogously, the metric tensor $g_{ij}(q)$ is assumed to be the direct sum $\bigoplus_{h=1}^N[g_{\mu\nu}(x_h)\oplus g_{\alpha,\beta}(y_h)]$ of the metric $g_{\mu\nu}(x_h)$ of spacetime and of the metric $g_{\alpha\beta}(y_h)$ of the group $G$.\\

With these premises, the connection between the CQG and the SQM is obtained by expanding the scalar Weyl wavefunction $\Psi(q)$ of the system as the harmonic superposition of products of the finite dimensional irreducible representations $g \to D^n(g)$ $(g \in G^N)$ of $G^N$~\cite{SalamStrathdee1982}, viz.
\begin{equation}\label{eq:Psifullexpansion}
  \Psi(q) = \Psi(x,y) = \sum_u\sum_{p,q=1}^u D^u[g^{-1}(y)]^p_{\;\;q}\psi^q_{\;\;pu}(x)
\end{equation}
where the sum includes all matrix elements of all finite-dimensional irreducible representations of the group $G^N$. The index $u$ fixes the dimensionality of the representations. Because the harmonics $D^u$ form a complete set of functions and the external fields $A_i(x,y)$ have been expressed in terms of the Killing vectors of $G^N$ (see Eq.~(\ref{eq:Ah})), insertion of Eq.~(\ref{eq:Psifullexpansion}) into Eq.~(\ref{eq:wave}) yields a set of wave equations for the expansion coefficients $\psi^q_{\;\;pu}(x)$ where only the spacetime potentials $A_\mu(x)$ and fields $F_m(x)$ appear explicitly. Each solution $\psi^q_{\;\;pu}(x)$ of these wave equations provides a corresponding solution to our fundamental Eqs.~(\ref{eq:HJ}), (\ref{eq:cont}), and (\ref{eq:wave}). In this way the problem posed by the coupled nonlinear Eqs.~(\ref{eq:HJ}) and (\ref{eq:cont}) is reduced to the problem of coupled linear wave equations in the spacetime coordinates $x$ only. The fields $\psi^q_{\;\;pu}(x)$ are not yet exactly the fields of the SQM, however. In fact, the expansion (\ref{eq:Psifullexpansion}) is too general and should be somewhat restricted to account for the conservation of the particle intrinsic helicity, as it will be discussed in the next Section where a single particle with spin is considered.
\section{The intrinsic helicity of a particle with spin}\label{sec:onespin}
In the case of a single particle with spin, the internal group $G$ is identified with the proper orthochronous Lorentz group $\tilde L_+$ so that the configuration space is $V=M_4\times \tilde L_+$ spanned by the coordinates $q^i=(x^\mu,y^\alpha)$. As internal coordinates we take the six Euler's angles $y^\alpha =(\alpha,\beta,\gamma,\varphi,\theta,\chi)$. Other parameterizations of the Lorentz group are possible, but Euler's angles are particularly suitable~\cite{DeMartiniSantamato2012,SantamatoDeMartini2013}. In fact, a direct calculation shows that the Killing vectors $K_\alpha^{\;\;m}(y)$ of $\tilde L_+$ are independent of the Euler angle $\gamma$ and so is the metric tensor $g_{\alpha\beta}(y)=K_\alpha^{\;\;m}(y)K_\beta^{\;\;n}(y)g_{mn}$, because the Killing metric of $\tilde L_+$ is given by $g_{mn}=\diag(1,1,1,-1,-1,-1)$. The external fields $A_i(q)=A_i(x,y)$ are given by $(e/c)(A_\mu(x),K_\alpha^{\;\;m}(y)F_m(x))$ [see Eq.~(\ref{eq:Ah})], where $A_\mu(x)=(\bm A(x),-\phi(x))$ are the e.m. potentials and $F_m(x)=(\bm H(x),\bm E(x))$ are the associated e.m. fields~\cite{DeMartiniSantamato2012,SantamatoDeMartini2013}. If we assume that $\rho(q)$ is independent of $\gamma$, it follows that the angle $\gamma$ never appears explicitly in Eq.~(\ref{eq:HJ}) and it can be considered as an ignorable (or cyclic) coordinate in the Hamilton-Jacobi equation of the CQG. We can then consider solutions $S(q)$ of Eq.~(\ref{eq:HJ}) of the form
\begin{equation}\label{eq:S0}
     S(q) = \hbar s \gamma + S_0(q)
\end{equation}
with constant $s$ and $S_0(q)$ independent of $\gamma$. Without loss of generality we may assume $s\ge 0$. The momentum $p_\gamma=\partial L^*/\partial\dot{\gamma}$ conjugate to the angle $\gamma$ is by definition the \qo{intrinsic helicity} of the particle. Equation~(\ref{eq:S0}) shows that along all trajectories of the complete figure introduced in Sec.~\ref{sec:stat} $p_\gamma=\partial_\gamma S$ takes the constant value $\hbar s$. Thus Eq.~(\ref{eq:S0}) reduces the accessible phase space associated to the Lagrangian $L^*(q,\dot{q})$ in Eq.~(\ref{eq:DeltaS}) to the region $p_\gamma = \hbar s$. We may also observe that the presence of the external fields $A_i$ does not affect the conservation of the value of $s$ so that we may consider the intrinsic helicity $s$ (in units of $\hbar$) as a fixed property of the particle, akin mass. According to Eq.~(\ref{eq:S0}), the scalar wavefunction $\Psi(q)$ given by the \textit{ansatz}~(\ref{eq:ansatz}) becomes
\begin{equation}\label{eq:PsitoPhi}
   \Psi_s(q) = e^{is\gamma}\Phi_s(q)
\end{equation}
with $\Phi_s(q)$ independent of $\gamma$. It can be easily verified that Eq.~(\ref{eq:PsitoPhi}) is consistent with the continuity Eq.~(\ref{eq:cont}) with $\rho(q)=|\Phi_s(q)|^2$ independent of $\gamma$, as assumed. \\

The harmonic expansion (\ref{eq:Psifullexpansion}) has not the factorized form required by Eq.~(\ref{eq:PsitoPhi}), in general. To be consistent with the conservation of the particle intrinsic helicity we must retain only the terms in the expansion~(\ref{eq:Psifullexpansion}) which are proportional to $e^{is\gamma}$ with given value of $s$. According to the methods of group theory~\cite{SalamStrathdee1982,KobayashiNomizu1996}, this selection should be made by restricting the harmonic expansion from the original group $G$ to the  homogeneous quotient space $G/H$ where $H$ is a subgroup of $G$. In our case $G$ is the proper Lorentz group $\tilde L_+$. To find $H$ we observe that $e^{is\gamma}\in U(1)$ provides the irreducible representation with helicity $s$ of the one-dimensional group $R_2$ ($\gamma$ is the group parameter) of the 2D rotations around a fixed axis, which is a subgroup of $\tilde L_+$. We then identify $H$ with $R_2$. The cosets of $\tilde L_+/R_2$ are labeled by a representative \qo{boost} $B(\tilde y)$~\cite{SalamStrathdee1982}, where $\tilde y$ denotes the set of Euler's angles which parameterize $\tilde L_+$ with $\gamma$ removed. The coordinates $\tilde y$ span the homogeneous space $\tilde L_+/R_2$. The form of the representative boost $B(\tilde y)$ is given once for ever and we have the so-called left translation rule $\bar\Lambda B(\tilde y)=B(\tilde y')R_2(\gamma')$ for fixed $\bar\Lambda\in \tilde L_+$~\cite{SalamStrathdee1982,KobayashiNomizu1996}. Both $\tilde y'$ and $\gamma'$ are determined by this relation as functions of $\bar\Lambda$ and $\tilde y$. The left translation rule implies the factorization $\Lambda(y)=B(\tilde y)R_2(\gamma)$ of an arbitrary Lorentz transformation $\Lambda$ into a boost $B(\tilde y)$ and a 2D rotation $R_2(\gamma)$\footnote{The boost $B(\tilde y)$ considered here differs from the boost commonly used in the relativistic mechanics because $B(\tilde y)$ contains a space rotation too. An alternative form of left translation rule is $R_2(\gamma')=B^{-1}(\tilde y')\bar\Lambda B(\tilde y)$, which represents \qo{Wigner's rotation} in the present case.}. In fact, we have $\bar\Lambda\Lambda=\bar\Lambda B(\tilde y)R_2(\gamma)=B(\tilde y')R_2(\gamma')R_2(\gamma)=B(\tilde y')R_2(\gamma'+\gamma)=\Lambda''$, which is the group property of $\tilde L_+$. Using this factorization, the function $\Phi_s(q)=\Phi_s(x,\tilde y)$ in Eq.~(\ref{eq:PsitoPhi}) is expanded in the harmonic series of the quotient space $\tilde L_+/R_2$, obtaining
\begin{equation}\label{eq:Phiexpansion}
  \Phi_s(q) = \Phi_s(x,\tilde y) = \sum_{p=-s}^s D^{(0,s)}[B^{-1}(\tilde y)]^s_{\;\;p}\psi^p(x)+
            \sum_{\dot p=-s}^s D^{(s,0)}[B^{-1}(\tilde y)]^s_{\;\;\dot p}\psi^{\dot p}(x)
\end{equation}
where $D^{(u,v)}[\Lambda]$ is the $(2u+1)\times (2v+1)$-dimensional representation of the Lorentz group. From the properties of the finite dimensional representations of the Lorentz group we see that $s$ is now restricted to have integer or half-integer non negative value, equal to the value of the quantum spin of the particle. Notice that both the undotted and dotted representations of $\tilde L_+$ are present in Eq.~(\ref{eq:Phiexpansion}), which preserves parity. The important case $s=1/2$ was worked out explicitly in Refs.~\cite{DeMartiniSantamato2012,SantamatoDeMartini2013} where it was shown that the four-component column vector $\psi_D(x) =\begin{pmatrix}
                                \psi^p(x) \\
                                \psi^{\dot p}(x) \\
                          \end{pmatrix}$
transforms as a Dirac spinor under Lorentz transformations and that the wave equation obeyed by $\Psi_D(x)$ was the square of Dirac's equation. In the light of this result for $s=1/2$, we assume that the Eqs.~(\ref{eq:PsitoPhi}) and (\ref{eq:Phiexpansion}) of the CQG describe the quantum spin for any other value integer or half-integer of $s$. By the methods of the SQM we can evaluate only the coefficients $\psi^q(x)$ of the expansion (\ref{eq:Phiexpansion}). Besides the spacetime coordinates $x^\mu$, the CQG includes additional coordinates $y^\alpha$ accounting for the internal degrees of freedom of the particle. As observed elsewhere, this renders the CQG a \textit{complete} theory in Einstein's sense, where the EPR and other paradoxes can be solved in spite of violation of Bell's inequalities~\cite{SantamatoDeMartini2012,DeMartiniSantamato2012a,DeMartiniSantamato2013,SantamatoDeMartini2013a,DeMartiniSantamato2014,DeMartiniSantamato2014b,DeMartiniSantamato2015}.    As final point we observe that the spinors of the SQM affect the harmonic expansion over the quotient space $\tilde L_+/R_2$ alone [see Eq.~(\ref{eq:Phiexpansion})], while the CQG includes also the global phase factor $e^{is\gamma}$ as shown in Eq.~(\ref{eq:PsitoPhi}). In other words, the SQM ignores the $\gamma$-rotation, which is accounted for by the CQG, instead. The effect of the $\gamma$ angle is to introduce a global phase factor in the scalar wavefunction $\Psi(q)$, which is neglected in the SQM, where only \textit{relative} phase have interest. But, as we shall see in the next Section, this global phase is really crucial in order to establish the correct Spin-Statistics connection, when identical particles are considered.
\section{The Spin Statistic Theorem}\label{sec:RSST}
Let us start with two identical particles first. The two-particle configuration space is the direct product $V^2 = V_1{\times}V_2$ of the configuration spaces of each particle. Let $q=(q_1,q_2)$ the coordinates $q$ of the two-particle configuration space $V^2$ with  $q_1=(x_1,y_1)$ and $q_2=(x_2,y_2)$. The spacetime positions $x_1$ and $x_2$ of the two particles are joined by a space-like geodesic, so that they can be simultaneously present to some inertial observer. The scalar wavefunction of the two-particle system is obtained by selecting the appropriate terms in the general expansion~(\ref{eq:Psifullexpansion}), viz.
\begin{eqnarray}\label{eq:Psi12}
   \Psi_s(q_1,q_2) &=& e^{is(\gamma_1+\gamma_2)}\left[\sum_{p_1,p_2=-s}^sD^{(0,s)}[B^{-1}(\tilde y_1)]^s_{\;\;p_1}D^{(0,s)}[B^{-1}(\tilde y_2)]^s_{\;\;p_2}
                       \Phi_{p_1p_2}(x_1,x_2) +  \nonumber \right.  \\
                       &&\left. \frac{}{} + \mathrm{\ dotted\ terms\ }\right]
\end{eqnarray}
where the omitted dotted terms have $D^{(s,0)}$ in place of $D^{(0,s)}$.\\

As said in Sec.~\ref{sec:stat}, singlevaluedness of the function $S(q_1, q_2) = S(x_1, y_1;x_2, y_2)$ is required to construct the Carath\'{e}odory complete figure and to gain a Weyl gauge-invariant statistical interpretation of quantum measurements. That $S(q_1, q_2)$ must be single valued is also evident from Eq.~(\ref{eq:DeltaS}): when the two particles are exchanged, their trajectories in the canonical bundle are also exchanged, which leaves unchanged the complete figure and the value of the integral (\ref{eq:DeltaS}). We require therefore the exchange symmetry $S(q_1,q_2) = S(q_2,q_1)$. Because $\rho(q_1,q_2) = \rho(q_2,q_1)$, that symmetry is equivalent to
\begin{equation}\label{eq:Psi12symm}
   \Psi(q_1,q_2) = \Psi(q_2,q_1)
\end{equation}
We notice -- once more -- that the scalar wavefunction $\Psi$ of the CQG is \textit{not} a wavefunction of the SQM (as we have shown, the wavefunctions of the SQM appear in the CQG as coefficients of suitable harmonic expansion of $\Psi$). Thus, the well known criticism against Bacry's and Broyles' argument~\cite{Bacry1995,Broyles1976} that the physical invariance of the of system under the exchange of identical particles would imply the same invariance of the wavefunction used by the SQM to describe the state of the system does not apply to the symmetry (\ref{eq:Psi12symm}).\\

From what discussed in the previous Sections it should be now evident that, among the coordinates $q^i$, the angle $\gamma$ requires special attention in the construction of the two-particle scalar wave function, because the phase factor $e^{is\gamma}$ leads to double-valued representations of the group $R_2(\gamma)$ when $s$ is half integer. The global phase factor $e^{is(\gamma_1+\gamma_2)}$ in Eq.~(\ref{eq:Psi12}) must be interpreted here as the irreducible unitary representation of the group $R_2(\gamma_1)\times R_2(\gamma_2)$. The separate conservation of the intrinsic helicity of the  two particles fixes this irreducible representation uniquely. Because $e^{is\gamma}$ and $e^{-is\gamma}$ are \textit{different} representations of $R_2(\gamma)$, it follows from $s>0$ that the angles $\gamma_1$ and $\gamma_2$ in Eq.~(\ref{eq:Psi12}) must be counted both counterclockwise. Therefore, we cannot  exchange $\gamma_1$ and $\gamma_2$ according to the natural rule $(\gamma_1\rightarrow\gamma_2,\gamma_2\rightarrow\gamma_1)$ because one of these two rules necessarily imply a forbidden clockwise change of $\gamma$, which belongs to a different representation of $R_2(\gamma)$. The correct way to exchange $\gamma_1$ and $\gamma_2$ without changing the group representation is $(\gamma_1\rightarrow\gamma_2,\gamma_2\rightarrow\gamma_1+2\pi)$ (we assumed here $\gamma_2 > \gamma_1$). The same conclusion about the right way to exchange $\gamma_1$ and $\gamma_2$ can be drawn by following different routes. For example, the space $R_2(\gamma_1)\times R_2(\gamma_2)$ can be replaced by the quotient space $(R_2(\gamma_1)\times R_2(\gamma_2))/S_2$, where $S_2$ is the symmetric group, with the singular points where $\gamma_1=\gamma_2$ removed~\cite{LeinaasMyrheim1977}. Alternatively, one can observe that counterclockwise and clockwise round trips over the circle belong to different homotopy classes so that choosing the anticlockwise direction is equivalent to chose one homotopy class~\cite{Jabs2010}. In our previous works we derived the exchange rule for the $\gamma$ angles from the \qo{ratchet} kinematical constraint $p_\gamma=\partial L^*/\partial\dot\gamma=\hbar s >0$, where $L^*$ is the Lagrange function in Eq.~(\ref{eq:DeltaS})~\cite{DeMartiniSantamato2014a,SantamatoDeMartini2015}. We think however that the group theoretical argument presented here is simpler and has the advantage to avoid the need to exchange the orientations of the frames attached to the two particles. No exchange restriction exists for the other coordinates, that can be exchanged using the natural rule. Having thus established the right exchange rules, we may apply them to Eq.~(\ref{eq:Psi12}) to obtain
\begin{eqnarray}\label{eq:Psi21}
   \Psi_s(q_2,q_1) &=& (-1)^{2s} e^{is(\gamma_1+\gamma_2)}\left[\sum_{p_1,p_2=-s}^s D^{(0,s)}[B^{-1}(\tilde y_2)]^s_{\;\;p_1}D^{(0,s)}[B^{-1}(\tilde y_1)]^s_{\;\;p_2}
                       \Phi_{p_1p_2}(x_2,x_1) +  \nonumber \right.  \\
                       &&\left. \frac{}{} + \mathrm{\ dotted\ terms\ }\right] = \nonumber \\
                   &=& (-1)^{2s} e^{is(\gamma_1+\gamma_2)}\left[\sum_{p_1,p_2=-s}^s D^{(0,s)}[B^{-1}(\tilde y_1)]^s_{\;\;p_1}D^{(0,s)}[B^{-1}(\tilde y_2)]^s_{\;\;p_2}
                       \Phi_{p_2p_1}(x_2,x_1) +  \nonumber \right.  \\
                       &&\left. \frac{}{} + \mathrm{\ dotted\ terms\ }\right]
\end{eqnarray}
where in the last term we exchanged the dummy indices $p_1$ and $p_2$. Now a direct comparison of Eq.~(\ref{eq:Psi21}) with Eq.~(\ref{eq:Psi12}) shows that the symmetry requirement (\ref{eq:Psi12symm}) can be satisfied only if
\begin{equation}\label{eq:Phi12symm}
  \Phi_{p_1p_2}(x_1,x_2) = (-1)^{2s}\Phi_{p_2p_1}(x_2,x_1).
\end{equation}
In other words, if we exchange \textit{both} the space-time coordinates $(x_1,x_2)$ \textit{and} the spin indices $(p_1,p_2)$ of the two particles, the space-time dependent part $\Phi_{p_1p_2}(x_1,x_2)$ of the scalar wavefunction (the only one considered in the SQM) must be chosen so to remain invariant when $s$ is integer and to change in sign when $s$ is half integer, in agreement with the Spin-Statistic connection\footnote{In the case $s=0$ we have $D^{(0,0)}=1$ and the SST reduces to the symmetry condition (\ref{eq:Psi12symm}).}. This result cannot be obtained so simply in the SQM, because the SQM ignores the $\gamma$ angle.\\

The proof of the SST is immediately extended to a system of $N$ identical particles. The requirement of  singlevaluedness of the action $S$ under arbitrary permutation of the particles leads to the symmetry
\begin{equation}\label{eq:Psisymm}
 \Psi(q_1,\dots,q_N) = \Psi(p(q_1,\dots,q_N))
\end{equation}
which generalizes Eq.~(\ref{eq:Psi12symm}). Because any permutation $p$ of the coordinates $q_h$ $(h=1,\dots,N)$ of the $N$ particles is obtained by a finite number $k_p$ of simple transpositions, we may apply the permutation $p$ to the scalar wavefunction $\Psi$ by applying $k_p$ exchanges of the coordinates of two particles at a time as described above. Each one of these exchanges inserts a phase factor $(-1)^{2s}$ to the scalar wavefunction so that, after the permutation $p$, $\Psi(q_1,\dots,q_N)$ is changed into $(-1)^{2k_p s}\Psi(p(q_1,\dots,q_N))$. Then, compatibility with the symmetry (\ref{eq:Psisymm}) implies that the space-time-dependent part of $\Psi(q_1,\dots,q_N)$ must satisfy the exchange symmetry
\begin{equation}\label{eq:Phisymm}
  \Phi_{p_1,\dots, p_N}(q_1,\dots,q_N) = (-1)^{2k_p s}\Phi_{p(p_1,\dots,p_N)}(p(q_1,\dots,q_N))
\end{equation}
which generalizes Eq.~(\ref{eq:Phi12symm}) and the Spin-Statistics connection to $N$ identical particles.Since $w(S)=0$ the symmetry requirement $S(q_1,q_2) = S(q_2,q_1)$ and the proof of the SST are Weyl gauge invariant as is due for meaningful physical laws.
\section{Conclusions}\label{sec:concl}
In conclusion, we presented a Weyl-gauge invariant derivation of the SST in curved spacetime and in the presence of external fields with no recourse to the quantum field approach. In the CQG framework, this is made possible by the peculiar role played by the conservation of the intrinsic helicity $p_\gamma$ of elementary particles and by the expansion properties of the scalar wavefunction $\Psi$ of the CQG in the harmonics of appropriate homogeneous space $G/H$. In the relativistic case the proof of the SST requires $G/H = \tilde L_+/R_2$, but the same proof can be carried out also in the nonrelativistic case by replacing $\tilde L_+/R_2$ with $R_3/R_2$, where $R_3$ is the group of 3D rotations~\cite{SantamatoDeMartini2015}. The proof presented here is somewhat simpler of the one in Ref.~\cite{SantamatoDeMartini2015}, because does not require the use of frames attached to the particle. The CQG handles bosons and fermions on the same footing by a unique scalar wave function $\Psi$ which remains unchanged in the particle permutation. It is the space-time dependent part $Phi$ of $\Psi$, the one containing the spinor fields of the SQM, which behaves differently under exchange of identical particles, according to Pauli's principle. The CQG derivation of the SST shows that Pauli's principle, as any other objective physical law, is Weyl-gauge invariant. Additional \qo{flavors} as isospin have been ignored and the metric tensor was held fixed. Both restrictions can be relaxed, in principle, and will be considered in the future. Nowadays the only physical effect due to the intrinsic helicity appears to be the \textquotedblleft Pauli's exclusion principle\textquotedblright, as shown here, but there is no reason of principle which forbids $p_\gamma$ to be measured, e.g. by a neutron  interferometry experiment. However, the realization of such experiment could be difficult, if not impossible, because no fields are known which actively interact with the intrinsic helicity of the elementary particles.


\begin{thebibliography}{55}%
\makeatletter
\providecommand \@ifxundefined [1]{%
 \@ifx{#1\undefined}
}%
\providecommand \@ifnum [1]{%
 \ifnum #1\expandafter \@firstoftwo
 \else \expandafter \@secondoftwo
 \fi
}%
\providecommand \@ifx [1]{%
 \ifx #1\expandafter \@firstoftwo
 \else \expandafter \@secondoftwo
 \fi
}%
\providecommand \natexlab [1]{#1}%
\providecommand \enquote  [1]{``#1''}%
\providecommand \bibnamefont  [1]{#1}%
\providecommand \bibfnamefont [1]{#1}%
\providecommand \citenamefont [1]{#1}%
\providecommand \href@noop [0]{\@secondoftwo}%
\providecommand \href [0]{\begingroup \@sanitize@url \@href}%
\providecommand \@href[1]{\@@startlink{#1}\@@href}%
\providecommand \@@href[1]{\endgroup#1\@@endlink}%
\providecommand \@sanitize@url [0]{\catcode `\\12\catcode `\$12\catcode
  `\&12\catcode `\#12\catcode `\^12\catcode `\_12\catcode `\%12\relax}%
\providecommand \@@startlink[1]{}%
\providecommand \@@endlink[0]{}%
\providecommand \url  [0]{\begingroup\@sanitize@url \@url }%
\providecommand \@url [1]{\endgroup\@href {#1}{\urlprefix }}%
\providecommand \urlprefix  [0]{URL }%
\providecommand \Eprint [0]{\href }%
\@ifxundefined \urlstyle {%
  \providecommand \doi  [0]{\begingroup \@sanitize@url \@doi}%
  \providecommand \@doi [1]{\endgroup \@@startlink {\doibase
  #1}doi:\discretionary {}{}{}#1\@@endlink }%
}{%
  \providecommand \doi  [0]{doi:\discretionary{}{}{}\begingroup
  \urlstyle{rm}\Url }%
}%
\providecommand \doibase [0]{http://dx.doi.org/}%
\providecommand \Doi [0]{\begingroup \@sanitize@url \@Doi }%
\providecommand \@Doi  [1]{\endgroup\@@startlink{\doibase#1}\@@Doi}%
\providecommand \@@Doi [1]{#1\@@endlink}%
\providecommand \selectlanguage [0]{\@gobble}%
\providecommand \bibinfo  [0]{\@secondoftwo}%
\providecommand \bibfield  [0]{\@secondoftwo}%
\providecommand \translation [1]{[#1]}%
\providecommand \BibitemOpen [0]{}%
\providecommand \bibitemStop [0]{}%
\providecommand \bibitemNoStop [0]{.\EOS\space}%
\providecommand \EOS [0]{\spacefactor3000\relax}%
\providecommand \BibitemShut  [1]{\csname bibitem#1\endcsname}%
\bibitem [{\citenamefont {Santamato}\ and\ \citenamefont
  {De~Martini}(2015)}]{SantamatoDeMartini2015}%
  \BibitemOpen
  \bibfield  {author} {\bibinfo {author} {\bibfnamefont {E.}~\bibnamefont
  {Santamato}}\ and\ \bibinfo {author} {\bibfnamefont {F.~D.}\ \bibnamefont
  {De~Martini}},\ }\Doi {10.1007/s10701-015-9912-7} {\bibfield  {journal}
  {\bibinfo  {journal} {Found. Phys.},\ }\textbf {\bibinfo {volume} {45}},\
  \bibinfo {pages} {858} (\bibinfo {year} {2015})},\ ISSN \bibinfo {issn}
  {0015-9018, 1572-9516}\BibitemShut {NoStop}%
\bibitem [{\citenamefont {Streater}\ and\ \citenamefont
  {Wightman}(2000)}]{StreaterWightman2000}%
  \BibitemOpen
  \bibfield  {author} {\bibinfo {author} {\bibfnamefont {R.~F.}\ \bibnamefont
  {Streater}}\ and\ \bibinfo {author} {\bibfnamefont {A.~S.}\ \bibnamefont
  {Wightman}},\ }\href@noop {} {\emph {\bibinfo {title} {PCT, Spin and
  Statistics, and All That}}}\ (\bibinfo  {publisher} {Princeton University
  Press},\ \bibinfo {address} {Princeton, N.J},\ \bibinfo {year} {2000})\ ISBN
  \bibinfo {isbn} {978-0-691-07062-9}\BibitemShut {NoStop}%
\bibitem [{\citenamefont {Fierz}(1939)}]{Fierz1939}%
  \BibitemOpen
  \bibfield  {author} {\bibinfo {author} {\bibfnamefont {M.}~\bibnamefont
  {Fierz}},\ }\href@noop {} {\bibfield  {journal} {\bibinfo  {journal} {Helv.
  Phys. Acta},\ }\textbf {\bibinfo {volume} {12}},\ \bibinfo {pages} {3}
  (\bibinfo {year} {1939})}\BibitemShut {NoStop}%
\bibitem [{\citenamefont {Pauli}(1940)}]{Pauli1940}%
  \BibitemOpen
  \bibfield  {author} {\bibinfo {author} {\bibfnamefont {W.}~\bibnamefont
  {Pauli}},\ }\Doi {10.1103/PhysRev.58.716} {\bibfield  {journal} {\bibinfo
  {journal} {Phys. Rev.},\ }\textbf {\bibinfo {volume} {58}},\ \bibinfo {pages}
  {716} (\bibinfo {year} {1940})},\ ISSN \bibinfo {issn} {0031-899X,
  1536-6065}\BibitemShut {NoStop}%
\bibitem [{\citenamefont {Schwinger}(1958)}]{Schwinger1958}%
  \BibitemOpen
  \bibfield  {author} {\bibinfo {author} {\bibfnamefont {J.}~\bibnamefont
  {Schwinger}},\ }\Doi {10.1073/pnas.44.2.223} {\bibfield  {journal} {\bibinfo
  {journal} {Proc. Natl. Acad. Sci. U.S.A.},\ }\textbf {\bibinfo {volume}
  {44}},\ \bibinfo {pages} {223} (\bibinfo {year} {1958})}\BibitemShut
  {NoStop}%
\bibitem [{\citenamefont {Wightman}(1956)}]{Wightman1956}%
  \BibitemOpen
  \bibfield  {author} {\bibinfo {author} {\bibfnamefont {A.~S.}\ \bibnamefont
  {Wightman}},\ }\Doi {10.1103/physrev.101.860} {\bibfield  {journal} {\bibinfo
   {journal} {Phys. Rev.},\ }\textbf {\bibinfo {volume} {101}},\ \bibinfo
  {pages} {860} (\bibinfo {year} {1956})}\BibitemShut {NoStop}%
\bibitem [{\citenamefont {Duck}\ \emph {et~al.}(1999)\citenamefont {Duck},
  \citenamefont {Sudarshan},\ and\ \citenamefont
  {Wightman}}]{DuckSudarshanWightman1999}%
  \BibitemOpen
  \bibfield  {author} {\bibinfo {author} {\bibfnamefont {I.}~\bibnamefont
  {Duck}}, \bibinfo {author} {\bibfnamefont {E.~C.~G.}\ \bibnamefont
  {Sudarshan}}, \ and\ \bibinfo {author} {\bibfnamefont {A.~S.}\ \bibnamefont
  {Wightman}},\ }\Doi {10.1119/1.19365} {\bibfield  {journal} {\bibinfo
  {journal} {Am. J. Phys},\ }\textbf {\bibinfo {volume} {67}},\ \bibinfo
  {pages} {742} (\bibinfo {year} {1999})},\ ISSN \bibinfo {issn} {0002-9505,
  1943-2909}\BibitemShut {NoStop}%
\bibitem [{\citenamefont {Duck}\ and\ \citenamefont
  {Sudarshan}(1998){\natexlab{a}}}]{DuckSudarshan1998a}%
  \BibitemOpen
  \bibfield  {author} {\bibinfo {author} {\bibfnamefont {I.}~\bibnamefont
  {Duck}}\ and\ \bibinfo {author} {\bibfnamefont {E.~C.~G.}\ \bibnamefont
  {Sudarshan}},\ }\href@noop {} {\emph {\bibinfo {title} {Pauli and the
  {Spin}-{Statistics} {Theorem}}}}\ (\bibinfo  {publisher} {World Scientific},\
  \bibinfo {year} {1998})\ ISBN \bibinfo {isbn} {978-981-4497-45-9}\BibitemShut
  {NoStop}%
\bibitem [{\citenamefont {Romer}(2002)}]{Romer2002}%
  \BibitemOpen
  \bibfield  {author} {\bibinfo {author} {\bibfnamefont {R.~H.}\ \bibnamefont
  {Romer}},\ }\Doi {10.1119/1.1482064} {\bibfield  {journal} {\bibinfo
  {journal} {Am. J. Phys},\ }\textbf {\bibinfo {volume} {70}},\ \bibinfo
  {pages} {791} (\bibinfo {year} {2002})},\ ISSN \bibinfo {issn}
  {00029505}\BibitemShut {NoStop}%
\bibitem [{\citenamefont {Duck}\ and\ \citenamefont
  {Sudarshan}(1998){\natexlab{b}}}]{DuckSudarshan1998}%
  \BibitemOpen
  \bibfield  {author} {\bibinfo {author} {\bibfnamefont {I.}~\bibnamefont
  {Duck}}\ and\ \bibinfo {author} {\bibfnamefont {E.~C.~G.}\ \bibnamefont
  {Sudarshan}},\ }\Doi {10.1119/1.18860} {\bibfield  {journal} {\bibinfo
  {journal} {Am. J. Phys},\ }\textbf {\bibinfo {volume} {66}},\ \bibinfo
  {pages} {284} (\bibinfo {year} {1998}{\natexlab{b}})},\ ISSN \bibinfo {issn}
  {0002-9505, 1943-2909}\BibitemShut {NoStop}%
\bibitem [{\citenamefont {Feynman}\ and\ \citenamefont
  {Leighton}(2011)}]{FeynmanLeighton2011}%
  \BibitemOpen
  \bibfield  {author} {\bibinfo {author} {\bibfnamefont {R.}~\bibnamefont
  {Feynman}}\ and\ \bibinfo {author} {\bibfnamefont {R.}~\bibnamefont
  {Leighton}},\ }\href@noop {} {\emph {\bibinfo {title} {Feynman {Lectures} on
  {Physics}, {Vol}. {I}: 1}}},\ \bibinfo {edition} {revised 50th anniversary}\
  ed.\ (\bibinfo  {publisher} {Basic Books},\ \bibinfo {address} {New York},\
  \bibinfo {year} {2011})\ ISBN \bibinfo {isbn} {978-0-465-02493-3}\BibitemShut
  {NoStop}%
\bibitem [{\citenamefont {Santamato}\ and\ \citenamefont {{De
  Martini}}(2013){\natexlab{a}}}]{SantamatoDeMartini2013a}%
  \BibitemOpen
  \bibfield  {author} {\bibinfo {author} {\bibfnamefont {E.}~\bibnamefont
  {Santamato}}\ and\ \bibinfo {author} {\bibfnamefont {F.}~\bibnamefont {{De
  Martini}}},\ }\Doi {10.1088/1742-6596/442/1/012059} {\bibfield  {journal}
  {\bibinfo  {journal} {J. Phys. Conf. Ser.},\ }\textbf {\bibinfo {volume}
  {442}},\ \bibinfo {pages} {012059} (\bibinfo {year} {2013}{\natexlab{a}})},\
  ISSN \bibinfo {issn} {1742-6596}\BibitemShut {NoStop}%
\bibitem [{\citenamefont {Santamato}\ and\ \citenamefont {{De
  Martini}}(2013){\natexlab{b}}}]{SantamatoDeMartini2013}%
  \BibitemOpen
  \bibfield  {author} {\bibinfo {author} {\bibfnamefont {E.}~\bibnamefont
  {Santamato}}\ and\ \bibinfo {author} {\bibfnamefont {F.}~\bibnamefont {{De
  Martini}}},\ }\Doi {10.1007/s10701-013-9703-y} {\bibfield  {journal}
  {\bibinfo  {journal} {Found. Phys.},\ }\textbf {\bibinfo {volume} {43}},\
  \bibinfo {pages} {631} (\bibinfo {year} {2013}{\natexlab{b}})}\BibitemShut
  {NoStop}%
\bibitem [{\citenamefont {Santamato}\ and\ \citenamefont {{De
  Martini}}(2012)}]{SantamatoDeMartini2012}%
  \BibitemOpen
  \bibfield  {author} {\bibinfo {author} {\bibfnamefont {E.}~\bibnamefont
  {Santamato}}\ and\ \bibinfo {author} {\bibfnamefont {F.}~\bibnamefont {{De
  Martini}}},\ }\Doi {10.1142/S0219749912410134} {\bibfield  {journal}
  {\bibinfo  {journal} {Int. J. Quantum Inf.},\ }\textbf {\bibinfo {volume}
  {10}},\ \bibinfo {pages} {1241013} (\bibinfo {year} {2012})},\ ISSN \bibinfo
  {issn} {0219-7499}\BibitemShut {NoStop}%
\bibitem [{\citenamefont {Santamato}(2008)}]{Santamato2008}%
  \BibitemOpen
  \bibfield  {author} {\bibinfo {author} {\bibfnamefont {E.}~\bibnamefont
  {Santamato}},\ }\href {http://arxiv.org/abs/0808.3237} {\bibfield  {journal}
  {\bibinfo  {journal} {arXiv:0808.3237 [quant-ph]}} (\bibinfo {year}
  {2008})},\ \bibinfo {note} {arXiv: 0808.3237}\BibitemShut {NoStop}%
\bibitem [{\citenamefont {{De Martini}}\ and\ \citenamefont
  {Santamato}(2015)}]{DeMartiniSantamato2015}%
  \BibitemOpen
  \bibfield  {author} {\bibinfo {author} {\bibfnamefont {F.}~\bibnamefont {{De
  Martini}}}\ and\ \bibinfo {author} {\bibfnamefont {E.}~\bibnamefont
  {Santamato}},\ }\Doi {10.1166/jap.2015.1195} {\bibfield  {journal} {\bibinfo
  {journal} {J. Adv. Phys.},\ }\textbf {\bibinfo {volume} {4}},\ \bibinfo
  {pages} {272} (\bibinfo {year} {2015})}\BibitemShut {NoStop}%
\bibitem [{\citenamefont {{De Martini}}\ and\ \citenamefont
  {Santamato}(2014){\natexlab{a}}}]{DeMartiniSantamato2014b}%
  \BibitemOpen
  \bibfield  {author} {\bibinfo {author} {\bibfnamefont {F.}~\bibnamefont {{De
  Martini}}}\ and\ \bibinfo {author} {\bibfnamefont {E.}~\bibnamefont
  {Santamato}},\ }\Doi {10.1088/0031-8949/2014/T163/014015} {\bibfield
  {journal} {\bibinfo  {journal} {Phys. Scr.},\ }\textbf {\bibinfo {volume}
  {T163}},\ \bibinfo {pages} {014015} (\bibinfo {year} {2014}{\natexlab{a}})},\
  ISSN \bibinfo {issn} {0031-8949, 1402-4896}\BibitemShut {NoStop}%
\bibitem [{\citenamefont {{De Martini}}\ and\ \citenamefont
  {Santamato}(2014){\natexlab{b}}}]{DeMartiniSantamato2014}%
  \BibitemOpen
  \bibfield  {author} {\bibinfo {author} {\bibfnamefont {F.}~\bibnamefont {{De
  Martini}}}\ and\ \bibinfo {author} {\bibfnamefont {E.}~\bibnamefont
  {Santamato}},\ }\Doi {10.1007/s10773-013-1651-y} {\bibfield  {journal}
  {\bibinfo  {journal} {Int. J. Theor. Phys.},\ }\textbf {\bibinfo {volume}
  {53}},\ \bibinfo {pages} {3308} (\bibinfo {year} {2014}{\natexlab{b}})},\
  ISSN \bibinfo {issn} {0020-7748, 1572-9575}\BibitemShut {NoStop}%
\bibitem [{\citenamefont {{De Martini}}\ and\ \citenamefont
  {Santamato}(2013)}]{DeMartiniSantamato2013}%
  \BibitemOpen
  \bibfield  {author} {\bibinfo {author} {\bibfnamefont {F.}~\bibnamefont {{De
  Martini}}}\ and\ \bibinfo {author} {\bibfnamefont {E.}~\bibnamefont
  {Santamato}},\ }\Doi {10.1051/epjconf/20135801012} {\bibfield  {journal}
  {\bibinfo  {journal} {EPJ Web Conf.},\ }\textbf {\bibinfo {volume} {58}},\
  \bibinfo {pages} {01012} (\bibinfo {year} {2013})},\ ISSN \bibinfo {issn}
  {2100-014X}\BibitemShut {NoStop}%
\bibitem [{\citenamefont {{De Martini}}\ and\ \citenamefont
  {Santamato}(2014){\natexlab{c}}}]{DeMartiniSantamato2014a}%
  \BibitemOpen
  \bibfield  {author} {\bibinfo {author} {\bibfnamefont {F.}~\bibnamefont {{De
  Martini}}}\ and\ \bibinfo {author} {\bibfnamefont {E.}~\bibnamefont
  {Santamato}},\ }\Doi {10.1142/S0219749915600047} {\bibfield  {journal}
  {\bibinfo  {journal} {Int. J. Quantum Inf.},\ }\textbf {\bibinfo {volume}
  {12}},\ \bibinfo {pages} {1560004} (\bibinfo {year} {2014}{\natexlab{c}})},\
  ISSN \bibinfo {issn} {0219-7499}\BibitemShut {NoStop}%
\bibitem [{\citenamefont {Weyl}(1952)}]{Weyl1952}%
  \BibitemOpen
  \bibfield  {author} {\bibinfo {author} {\bibfnamefont {H.}~\bibnamefont
  {Weyl}},\ }\href@noop {} {\emph {\bibinfo {title} {Space, {Time},
  {Matter}}}},\ \bibinfo {edition} {4th}\ ed.\ (\bibinfo  {publisher} {Dover
  Publications, Inc.},\ \bibinfo {address} {New York},\ \bibinfo {year}
  {1952})\BibitemShut {NoStop}%
\bibitem [{Note1()}]{Note1}%
  \BibitemOpen
  \bibinfo {note} {In some axiomatic approaches to the structure of spacetime,
  the so-called \textquotedblleft Weyl tensor\textquotedblright $W^i_{jkl}$ is
  used in place of the curvature tensor $R^i_{jkl}$. The Weyl tensor is
  obtained from the full curvature tensor by subtracting out various traces and
  describes the tidal part of the gravitational forces. All contractions of
  $W^i_{jkl}$ are zero.}\BibitemShut {Stop}%
\bibitem [{\citenamefont {Weyl}(1918)}]{Weyl1918}%
  \BibitemOpen
  \bibfield  {author} {\bibinfo {author} {\bibfnamefont {H.}~\bibnamefont
  {Weyl}},\ }\href@noop {} {\bibfield  {journal} {\bibinfo  {journal} {Sitz.
  Berichte d. Preuss. Akad. d. Wiss. Berlin},\ }\textbf {\bibinfo {volume}
  {K1}},\ \bibinfo {pages} {465} (\bibinfo {year} {1918})},\ \bibinfo {note}
  {reprinted in: \textit{The Principles of Relativity} (Dover, New York,
  1923)}\BibitemShut {NoStop}%
\bibitem [{\citenamefont {Einstein}(1918)}]{einstein1918}%
  \BibitemOpen
  \bibfield  {author} {\bibinfo {author} {\bibfnamefont {A.}~\bibnamefont
  {Einstein}},\ }\href@noop {} {\bibfield  {journal} {\bibinfo  {journal}
  {Sitzung. d. Preuss. Akad. d. Wiss},\ }\textbf {\bibinfo {volume} {K1}},\
  \bibinfo {pages} {478} (\bibinfo {year} {1918})},\ \bibinfo {note} {including
  Weyl's reply.}\BibitemShut {Stop}%
\bibitem [{\citenamefont {Dirac}(1973)}]{Dirac1973}%
  \BibitemOpen
  \bibfield  {author} {\bibinfo {author} {\bibfnamefont {P.~A.~M.}\
  \bibnamefont {Dirac}},\ }\href@noop {} {\bibfield  {journal} {\bibinfo
  {journal} {Proc. R. Soc. London, Ser. A},\ }\textbf {\bibinfo {volume}
  {333}},\ \bibinfo {pages} {403} (\bibinfo {year} {1973})}\BibitemShut
  {NoStop}%
\bibitem [{\citenamefont {Utiyama}(1973)}]{Utiyama1973}%
  \BibitemOpen
  \bibfield  {author} {\bibinfo {author} {\bibfnamefont {R.}~\bibnamefont
  {Utiyama}},\ }\href@noop {} {\bibfield  {journal} {\bibinfo  {journal} {Prog.
  Theor. Phys},\ }\textbf {\bibinfo {volume} {50}},\ \bibinfo {pages} {2080}
  (\bibinfo {year} {1973})}\BibitemShut {NoStop}%
\bibitem [{\citenamefont {Utiyama}(1975)}]{Utiyama1975}%
  \BibitemOpen
  \bibfield  {author} {\bibinfo {author} {\bibfnamefont {R.}~\bibnamefont
  {Utiyama}},\ }\href@noop {} {\bibfield  {journal} {\bibinfo  {journal} {Prog.
  Theor. Phys},\ }\textbf {\bibinfo {volume} {53}},\ \bibinfo {pages} {565}
  (\bibinfo {year} {1975})}\BibitemShut {NoStop}%
\bibitem [{\citenamefont {Cheng}(1988)}]{Cheng1988}%
  \BibitemOpen
  \bibfield  {author} {\bibinfo {author} {\bibfnamefont {H.}~\bibnamefont
  {Cheng}},\ }\href@noop {} {\bibfield  {journal} {\bibinfo  {journal} {Phys.
  Rev. Lett.},\ }\textbf {\bibinfo {volume} {61}},\ \bibinfo {pages} {2182}
  (\bibinfo {year} {1988})}\BibitemShut {NoStop}%
\bibitem [{\citenamefont {Hayashi}\ \emph {et~al.}(1977)\citenamefont
  {Hayashi}, \citenamefont {Kasuya},\ and\ \citenamefont
  {Shirafuji}}]{HayashiKasuyaShirafuji1977}%
  \BibitemOpen
  \bibfield  {author} {\bibinfo {author} {\bibfnamefont {K.}~\bibnamefont
  {Hayashi}}, \bibinfo {author} {\bibfnamefont {M.}~\bibnamefont {Kasuya}}, \
  and\ \bibinfo {author} {\bibfnamefont {T.}~\bibnamefont {Shirafuji}},\
  }\href@noop {} {\bibfield  {journal} {\bibinfo  {journal} {Prog. Theor.
  Phys},\ }\textbf {\bibinfo {volume} {57}},\ \bibinfo {pages} {431} (\bibinfo
  {year} {1977})}\BibitemShut {NoStop}%
\bibitem [{\citenamefont {Wheeler}(1990)}]{Wheeler1990}%
  \BibitemOpen
  \bibfield  {author} {\bibinfo {author} {\bibfnamefont {J.~T.}\ \bibnamefont
  {Wheeler}},\ }\href@noop {} {\bibfield  {journal} {\bibinfo  {journal} {Phys.
  Rev. D},\ }\textbf {\bibinfo {volume} {41}},\ \bibinfo {pages} {431}
  (\bibinfo {year} {1990})}\BibitemShut {NoStop}%
\bibitem [{\citenamefont {Wheeler}(1998)}]{Wheeler1998}%
  \BibitemOpen
  \bibfield  {author} {\bibinfo {author} {\bibfnamefont {J.~T.}\ \bibnamefont
  {Wheeler}},\ }\href@noop {} {\bibfield  {journal} {\bibinfo  {journal} {J.
  Math. Phys.},\ }\textbf {\bibinfo {volume} {39}},\ \bibinfo {pages} {299}
  (\bibinfo {year} {1998})}\BibitemShut {NoStop}%
\bibitem [{\citenamefont {Ehlers}\ \emph {et~al.}(1972)\citenamefont {Ehlers},
  \citenamefont {Pirani},\ and\ \citenamefont {Schild}}]{J.EhlersA.Schild1972}%
  \BibitemOpen
  \bibfield  {author} {\bibinfo {author} {\bibfnamefont {J.}~\bibnamefont
  {Ehlers}}, \bibinfo {author} {\bibfnamefont {F.~A.}\ \bibnamefont {Pirani}},
  \ and\ \bibinfo {author} {\bibfnamefont {A.}~\bibnamefont {Schild}},\ }in\
  \href@noop {} {\emph {\bibinfo {booktitle} {General Relativity (Papers in
  honour of J. L. Synge)}}},\ \bibinfo {editor} {edited by\ \bibinfo {editor}
  {\bibfnamefont {L.}~\bibnamefont {O'Raifeartaigh}}}\ (\bibinfo  {publisher}
  {Clarendon Press, Oxford},\ \bibinfo {year} {1972})\ p.~\bibinfo {pages}
  {63}\BibitemShut {NoStop}%
\bibitem [{\citenamefont {Hochberg}\ and\ \citenamefont
  {Plunien}(1991)}]{HochbergPlunien1991}%
  \BibitemOpen
  \bibfield  {author} {\bibinfo {author} {\bibfnamefont {D.}~\bibnamefont
  {Hochberg}}\ and\ \bibinfo {author} {\bibfnamefont {G.}~\bibnamefont
  {Plunien}},\ }\Doi {10.1103/PhysRevD.43.3358} {\bibfield  {journal} {\bibinfo
   {journal} {Phys. Rev. D},\ }\textbf {\bibinfo {volume} {43}},\ \bibinfo
  {pages} {3358} (\bibinfo {year} {1991})}\BibitemShut {NoStop}%
\bibitem [{\citenamefont {Faraoni}\ and\ \citenamefont
  {Capozziello}(2011)}]{faraoni_beyond_2011}%
  \BibitemOpen
  \bibfield  {author} {\bibinfo {author} {\bibfnamefont {V.}~\bibnamefont
  {Faraoni}}\ and\ \bibinfo {author} {\bibfnamefont {S.}~\bibnamefont
  {Capozziello}},\ }\href {http://link.springer.com/10.1007/978-94-007-0165-6}
  {\emph {\bibinfo {title} {Beyond {Einstein} {Gravity}}}}\ (\bibinfo
  {publisher} {Springer Netherlands},\ \bibinfo {address} {Dordrecht},\
  \bibinfo {year} {2011})\ ISBN \bibinfo {isbn} {978-94-007-0164-9
  978-94-007-0165-6}\BibitemShut {NoStop}%
\bibitem [{\citenamefont {Trautman}(2012)}]{Trautman2012}%
  \BibitemOpen
  \bibfield  {author} {\bibinfo {author} {\bibfnamefont {A.}~\bibnamefont
  {Trautman}},\ }\href@noop {} {\bibfield  {journal} {\bibinfo  {journal} {Gen.
  Relativ. Gravitation},\ }\textbf {\bibinfo {volume} {44}},\ \bibinfo {pages}
  {1581} (\bibinfo {year} {2012})}\BibitemShut {NoStop}%
\bibitem [{\citenamefont {Fatibene}\ \emph {et~al.}(2015)\citenamefont
  {Fatibene}, \citenamefont {Garruto},\ and\ \citenamefont
  {Polistina}}]{FatibeneGarrutoPolistina2015}%
  \BibitemOpen
  \bibfield  {author} {\bibinfo {author} {\bibfnamefont {L.}~\bibnamefont
  {Fatibene}}, \bibinfo {author} {\bibfnamefont {S.}~\bibnamefont {Garruto}}, \
  and\ \bibinfo {author} {\bibfnamefont {M.}~\bibnamefont {Polistina}},\ }\Doi
  {10.1142/s0219887815500449} {\bibfield  {journal} {\bibinfo  {journal} {Int.
  J. Geom. Meth. Mod. Phys.},\ }\textbf {\bibinfo {volume} {12}},\ \bibinfo
  {pages} {1550044} (\bibinfo {year} {2015})}\BibitemShut {NoStop}%
\bibitem [{\citenamefont {Santamato}(1984){\natexlab{a}}}]{Santamato1984}%
  \BibitemOpen
  \bibfield  {author} {\bibinfo {author} {\bibfnamefont {E.}~\bibnamefont
  {Santamato}},\ }\Doi {10.1103/PhysRevD.29.216} {\bibfield  {journal}
  {\bibinfo  {journal} {Phys. Rev. D},\ }\textbf {\bibinfo {volume} {29}},\
  \bibinfo {pages} {216} (\bibinfo {year} {1984}{\natexlab{a}})}\BibitemShut
  {NoStop}%
\bibitem [{\citenamefont {Santamato}(1984){\natexlab{b}}}]{Santamato1984a}%
  \BibitemOpen
  \bibfield  {author} {\bibinfo {author} {\bibfnamefont {E.}~\bibnamefont
  {Santamato}},\ }\Doi {10.1063/1.526467} {\bibfield  {journal} {\bibinfo
  {journal} {J. Math. Phys.},\ }\textbf {\bibinfo {volume} {25}},\ \bibinfo
  {pages} {2477} (\bibinfo {year} {1984}{\natexlab{b}})},\ ISSN \bibinfo {issn}
  {0022-2488, 1089-7658}\BibitemShut {NoStop}%
\bibitem [{\citenamefont {Santamato}(1985)}]{Santamato1985}%
  \BibitemOpen
  \bibfield  {author} {\bibinfo {author} {\bibfnamefont {E.}~\bibnamefont
  {Santamato}},\ }\Doi {10.1103/PhysRevD.32.2615} {\bibfield  {journal}
  {\bibinfo  {journal} {Phys. Rev. D},\ }\textbf {\bibinfo {volume} {32}},\
  \bibinfo {pages} {2615} (\bibinfo {year} {1985})}\BibitemShut {NoStop}%
\bibitem [{\citenamefont {Santamato}(1988)}]{Santamato1988}%
  \BibitemOpen
  \bibfield  {author} {\bibinfo {author} {\bibfnamefont {E.}~\bibnamefont
  {Santamato}},\ }\Doi {10.1016/0375-9601(88)90593-2} {\bibfield  {journal}
  {\bibinfo  {journal} {Phys. Lett. A},\ }\textbf {\bibinfo {volume} {130}},\
  \bibinfo {pages} {199} (\bibinfo {year} {1988})},\ ISSN \bibinfo {issn}
  {0375-9601}\BibitemShut {NoStop}%
\bibitem [{\citenamefont {Bohm}(1952){\natexlab{a}}}]{Bohm1952}%
  \BibitemOpen
  \bibfield  {author} {\bibinfo {author} {\bibfnamefont {D.}~\bibnamefont
  {Bohm}},\ }\Doi {10.1103/PhysRev.85.166} {\bibfield  {journal} {\bibinfo
  {journal} {Phys. Rev.},\ }\textbf {\bibinfo {volume} {85}},\ \bibinfo {pages}
  {166} (\bibinfo {year} {1952}{\natexlab{a}})}\BibitemShut {NoStop}%
\bibitem [{\citenamefont {Bohm}(1952){\natexlab{b}}}]{Bohm1952a}%
  \BibitemOpen
  \bibfield  {author} {\bibinfo {author} {\bibfnamefont {D.}~\bibnamefont
  {Bohm}},\ }\Doi {10.1103/PhysRev.85.180} {\bibfield  {journal} {\bibinfo
  {journal} {Phys. Rev.},\ }\textbf {\bibinfo {volume} {85}},\ \bibinfo {pages}
  {180} (\bibinfo {year} {1952}{\natexlab{b}})}\BibitemShut {NoStop}%
\bibitem [{\citenamefont {Bohm}\ and\ \citenamefont
  {Hiley}(1995)}]{BohmHiley1995}%
  \BibitemOpen
  \bibfield  {author} {\bibinfo {author} {\bibfnamefont {D.}~\bibnamefont
  {Bohm}}\ and\ \bibinfo {author} {\bibfnamefont {B.~J.}\ \bibnamefont
  {Hiley}},\ }\href@noop {} {\emph {\bibinfo {title} {The {Undivided}
  {Universe}: {An} {Ontological} {Interpretation} of {Quantum} {Theory}}}},\
  \bibinfo {edition} {reprint}\ ed.\ (\bibinfo  {publisher} {Routledge},\
  \bibinfo {address} {London; New York},\ \bibinfo {year} {1995})\ ISBN
  \bibinfo {isbn} {978-0-415-12185-9}\BibitemShut {NoStop}%
\bibitem [{\citenamefont {Rund}(1973)}]{Rund1973}%
  \BibitemOpen
  \bibfield  {author} {\bibinfo {author} {\bibfnamefont {H.}~\bibnamefont
  {Rund}},\ }\href@noop {} {\emph {\bibinfo {title} {The Hamilton-{Jacobi}
  {Theory} in the {Calculus} of {Variations} {Its} {Role} in {Mathematics}
  {Theory} and {Application}}}}\ (\bibinfo  {publisher} {Krieger Pub Co},\
  \bibinfo {address} {Huntington, N.Y.},\ \bibinfo {year} {1973})\ ISBN
  \bibinfo {isbn} {978-0-88275-063-7}\BibitemShut {NoStop}%
\bibitem [{\citenamefont {Ballentine}(1970)}]{Ballentine1970}%
  \BibitemOpen
  \bibfield  {author} {\bibinfo {author} {\bibfnamefont {L.~E.}\ \bibnamefont
  {Ballentine}},\ }\href@noop {} {\bibfield  {journal} {\bibinfo  {journal}
  {Rev. Mod. Phys.},\ }\textbf {\bibinfo {volume} {42}},\ \bibinfo {pages}
  {358} (\bibinfo {year} {1970})}\BibitemShut {NoStop}%
\bibitem [{\citenamefont {De~Martini}\ and\ \citenamefont
  {Santamato}(2012)}]{DeMartiniSantamato2012}%
  \BibitemOpen
  \bibfield  {author} {\bibinfo {author} {\bibfnamefont {F.}~\bibnamefont
  {De~Martini}}\ and\ \bibinfo {author} {\bibfnamefont {E.}~\bibnamefont
  {Santamato}},\ }in\ \Doi {10.1063/1.3688951} {\emph {\bibinfo {booktitle}
  {Foundations of {Probability} and {Physics} 6}}},\ \bibinfo {series} {{AIP}
  {Conference} {Proceedings}}, Vol.~\bibinfo {volume} {4}\ (\bibinfo
  {publisher} {D'Ariano, Mauro and Fei, Shao-Ming and Haven, Emmanuel and
  Hiesmayr, Beatrix and Jaeger, Gregg and Khrennikov, Andrei and Larsson,
  Jan-Ake},\ \bibinfo {address} {Melville},\ \bibinfo {year} {2012})\ pp.\
  \bibinfo {pages} {45--54}\BibitemShut {NoStop}%
\bibitem [{\citenamefont {{De Martini}}\ and\ \citenamefont
  {Santamato}(2012)}]{DeMartiniSantamato2012a}%
  \BibitemOpen
  \bibfield  {author} {\bibinfo {author} {\bibfnamefont {F.}~\bibnamefont {{De
  Martini}}}\ and\ \bibinfo {author} {\bibfnamefont {E.}~\bibnamefont
  {Santamato}}\ }(\bibinfo  {publisher} {Khrennikov, Andrei and Atmanspacher,
  H. and Migdall, Alan and Polyakov, SergeyAIP Conference Proceedings},\
  \bibinfo {year} {2012})\ pp.\ \bibinfo {pages} {162--171}\BibitemShut
  {NoStop}%
\bibitem [{\citenamefont {Salam}\ and\ \citenamefont
  {Strathdee}(1982)}]{SalamStrathdee1982}%
  \BibitemOpen
  \bibfield  {author} {\bibinfo {author} {\bibfnamefont {A.}~\bibnamefont
  {Salam}}\ and\ \bibinfo {author} {\bibfnamefont {J.}~\bibnamefont
  {Strathdee}},\ }\Doi {10.1016/0003-4916(82)90291-3} {\bibfield  {journal}
  {\bibinfo  {journal} {Ann. Phys.},\ }\textbf {\bibinfo {volume} {141}},\
  \bibinfo {pages} {316} (\bibinfo {year} {1982})},\ ISSN \bibinfo {issn}
  {0003-4916}\BibitemShut {NoStop}%
\bibitem [{\citenamefont {Kobayashi}\ and\ \citenamefont
  {Nomizu}(1996)}]{KobayashiNomizu1996}%
  \BibitemOpen
  \bibfield  {author} {\bibinfo {author} {\bibfnamefont {S.}~\bibnamefont
  {Kobayashi}}\ and\ \bibinfo {author} {\bibfnamefont {K.}~\bibnamefont
  {Nomizu}},\ }\href@noop {} {\emph {\bibinfo {title} {Foundations of
  {Differential} {Geometry}, {Vol}. 2}}},\ \bibinfo {edition} {volume 2
  edition}\ ed.\ (\bibinfo  {publisher} {Wiley-Interscience},\ \bibinfo
  {address} {New York},\ \bibinfo {year} {1996})\ ISBN \bibinfo {isbn}
  {978-0-471-15732-8}\BibitemShut {NoStop}%
\bibitem [{Note2()}]{Note2}%
  \BibitemOpen
  \bibinfo {note} {The boost $B(\protect \mathaccentV {tilde}07Ey)$ considered
  here differs from the boost commonly used in the relativistic mechanics
  because it contains a space rotation too.}\BibitemShut {Stop}%
\bibitem [{\citenamefont {Bacry}(1995)}]{Bacry1995}%
  \BibitemOpen
  \bibfield  {author} {\bibinfo {author} {\bibfnamefont {H.}~\bibnamefont
  {Bacry}},\ }\Doi {10.1119/1.17952} {\bibfield  {journal} {\bibinfo  {journal}
  {Am. J. Phys},\ }\textbf {\bibinfo {volume} {63}},\ \bibinfo {pages} {297}
  (\bibinfo {year} {1995})},\ ISSN \bibinfo {issn} {0002-9505,
  1943-2909}\BibitemShut {NoStop}%
\bibitem [{\citenamefont {Broyles}(1976)}]{Broyles1976}%
  \BibitemOpen
  \bibfield  {author} {\bibinfo {author} {\bibfnamefont {A.~A.}\ \bibnamefont
  {Broyles}},\ }\Doi {10.1119/1.10191} {\bibfield  {journal} {\bibinfo
  {journal} {Am. J. Phys},\ }\textbf {\bibinfo {volume} {44}},\ \bibinfo
  {pages} {340} (\bibinfo {year} {1976})},\ ISSN \bibinfo {issn} {0002-9505,
  1943-2909}\BibitemShut {NoStop}%
\bibitem [{\citenamefont {Leinaas}\ and\ \citenamefont
  {Myrheim}(1977)}]{LeinaasMyrheim1977}%
  \BibitemOpen
  \bibfield  {author} {\bibinfo {author} {\bibfnamefont {J.~M.}\ \bibnamefont
  {Leinaas}}\ and\ \bibinfo {author} {\bibfnamefont {J.}~\bibnamefont
  {Myrheim}},\ }\Doi {10.1007/BF02727953} {\bibfield  {journal} {\bibinfo
  {journal} {Nuovo Cimento B Series 11},\ }\textbf {\bibinfo {volume} {37}},\
  \bibinfo {pages} {1} (\bibinfo {year} {1977})},\ ISSN \bibinfo {issn}
  {0369-3554, 1826-9877}\BibitemShut {NoStop}%
\bibitem [{\citenamefont {Jabs}(2010)}]{Jabs2010}%
  \BibitemOpen
  \bibfield  {author} {\bibinfo {author} {\bibfnamefont {A.}~\bibnamefont
  {Jabs}},\ }\href@noop {} {\bibfield  {journal} {\bibinfo  {journal} {Found.
  Phys.},\ }\textbf {\bibinfo {volume} {40}},\ \bibinfo {pages} {776} (\bibinfo
  {year} {2010})}\BibitemShut {NoStop}%
\end{thebibliography}
%
\end{document}